# DETECTING COMMONALITY AND VARIABILITY IN USE-CASE DIAGRAM VARIANTS

[1]RA'FAT AL-MSIE'DEEN, [2]ANAS H. BLASI, [1]HAMZEH EYAL SALMAN, [3]SAQER S. ALJA'AFREH, [4]AHMAD ABADLEH, [5]MOHAMMED A. ALSUWAIKET, [2]AWNI HAMMOURI, [1]ASMAA JAMEEL AL_NAWAISEH, [1]WAFA TARAWNEH, [1]SULEYMAN A. AL-SHOWARAH

[1]Department of Software Engineering, Faculty of IT, Mutah University, Mutah 61710, Karak, Jordan

[2]Department of Data Science, Faculty of IT, Mutah University, Mutah 61710, Karak, Jordan

[3]Department of Electrical Engineering, Faculty of Engineering, Mutah University, Karak, Jordan

[4]Department of Computer Science, Faculty of IT, Mutah University, Mutah 61710, Karak, Jordan

[5]Department of CS and Engineering Technology, Hafar Batin University, Hafar Batin, Saudi Arabia

E-mail: [1]rafatalmsiedeen@mutah.edu.jo, [2]ablasi1@mutah.edu.jo, [1]hamzehmu@mutah.edu.jo, [3]eng.saqer-jaa@mutah.edu.jo, [4]ahmad_a@mutah.edu.jo, [5]malsuwaiket@uhb.edu.sa, [2]hammouri@mutah.edu.jo, [1]asma@mutah.edu.jo, [1]wafa@mutah.edu.jo, [1]showarah@mutah.edu.jo

## ABSTRACT

The use-case diagram is a software artifact. Thus, as with any software artifact, the use-case diagrams change across time through the software development life cycle. Therefore, several versions of the same diagram are existed at distinct times. Thus, comparing all use-case diagram variants to detect common and variable use-cases becomes one of the main challenges in the product line reengineering field. The contribution of this paper is to suggest an automatic approach to compare a collection of use-case diagram variants and detect both commonality and variability. In our work, every use-case represents a feature. The proposed approach visualizes the detected features using formal concept analysis, where common and variable features are introduced to software engineers. The proposed approach was applied on a mobile media case study to be validated. The findings confirm the importance and the performance of the suggested approach as all common and variable features were precisely detected via formal concept analysis and latent semantic indexing.

## 1. INTRODUCTION

The challenge of comparing model or diagram variants is well identified in Software Product Line (SPL) reengineering [1], [2]. However, the main issue is to analyze a collection of model or diagram variants to detect commonality and variability [3]. In our work, use-case diagram variants are typically characterized by two sets of use-cases: the use-cases that are shared by all use-case diagram variants, represent the SPL's commonalities, and the use-cases that are shared by some but not all use-case diagram variants, represent the SPL's variability. Feature Model (FM) is a hierarchical form depicting commonalities and variabilities in SPL [4], [5]. Figure 1 illustrates an example of a FM describing mobile media software variants.

Software variants frequently change from an original product created for and effectively used by the first client. Mobile media [6] is one of the several examples of such product evolution [7]. Use-case diagram variants usually share some common use-cases, but they are also different from one to another due to subsequent customization to meet particular requirements of various clients [8].

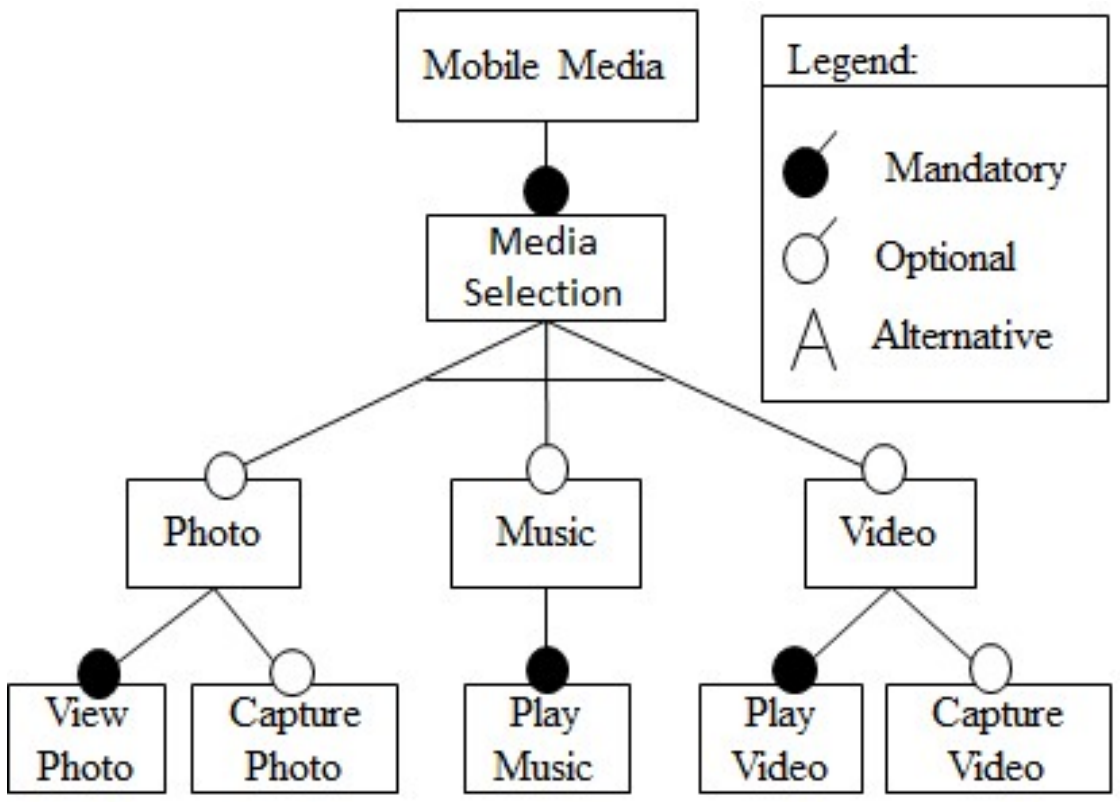

*Figure 1: FM of mobile media variants (partial).*

This paper is suggested an approach to detect commonality and variability from use-case diagrams for a group of software variants. The first step is to define the FM of SPL to build. Our approach accepts as inputs the use-case diagram variants. All use-cases form the initial search space. We rely on Formal Concept Analysis (or FCA for short) to decrease this search space by distinguishing common and variable use-cases, then separating the set of variable use-cases into more reduced sub-groups. For the minimal subgroups, we use Latent Semantic Indexing (LSI) to find the similarity among use-case and their description. Then, we use FCA again to identify common and variable use-cases based on LSI results.

In this paper, we rely on FCA to obtain an ordered set of concepts from a dataset (named a Formal Context) created of objects expressed by attributes. Information Retrieval (IR) [9] has been recognized as useful in different fields like software maintenance and evolution. In the proposed case, LSI has been used to find the lexical similarity among use-cases and their description [10], [11]. The efficacy of IR techniques is determined using IR metrics: precision, recall, and F-Measure [12], [13], [14]. Use-case diagrams assist the modeling of functional variability. Also, use-case diagrams can be utilized to define common and variable behavioral characteristics of software variants [15]. In this study, we assume that each use-case represents a functional feature, and we consider only the use-cases without actors and use-case relationships (*i.e.*, extend and include). Figure 2 displays an example of a use-case diagram of the first release of mobile media software [16]. The interested reader can get extra information about FCA, LSI, and use-case diagram in [17], [18], [19].

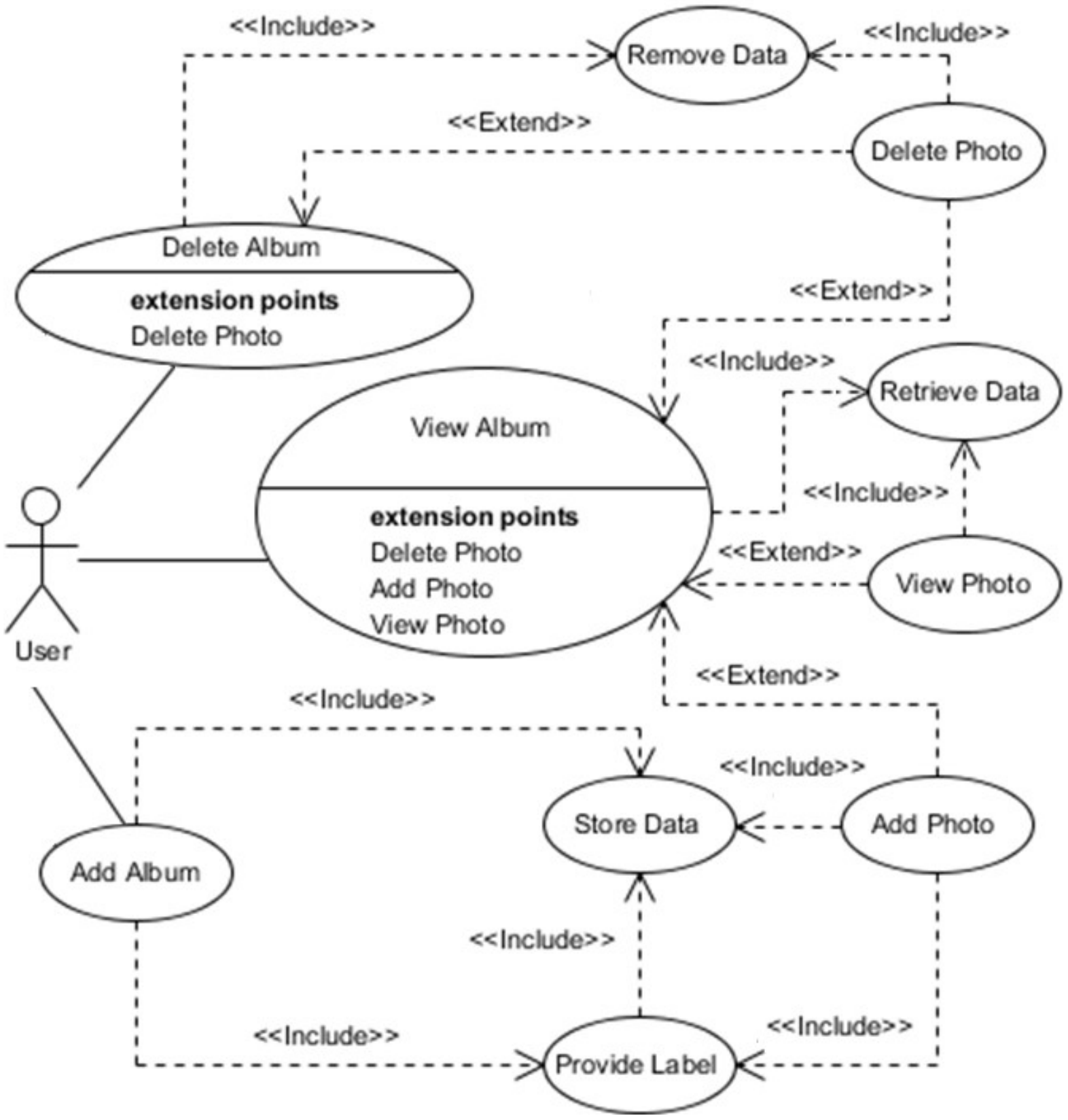

*Figure 2: Use-case diagram of mobile media release 1.*

The proposed method is described in the rest of this paper as follows. Section 2 introduces the mobile media variants. Section 3 offers an overview of the suggested method. Section 4 describes the detection process of commonality and variability across use-case diagram variants. Section 5 presents the experimental results. Section 6 presents the related work closest to our method. Finally, section 7 gives the conclusion and perspectives.

## 2. MOBILE MEDIA VARIANTS

Mobile media is a software product variant constructed to allow the end user of a mobile device to execute several choices (*e.g.,* playing music or videos). The system has been constructed as variants in eight various releases [6]. Each of these releases combines some distinct features to the system. For instance, release one implements the initial product with only the functionality of seeing photos and arranging them by albums. Release 2 and 3 include error handling and the implementation of several variable features (*e.g.,* edit labels). Table 1 shows the evolution across product variants of mobile media.

*Table 1: Mobile media product variants.*

| Release | Description |
|---|---|
| **r1** | Mobile media core |
| **r2** | Error handling included |
| **r3** | Sort media by frequency and edit caption service added |
| **r4** | Set favourites photos added |
| **r5** | Added a service for copy media to another album |
| **r6** | Added a service for sending media by SMS |
| **r7** | Added the service for playing media |
| **r8** | Added the service for playing media and capture media |

Our proposed approach takes use-case diagrams of a set of variants and their use-case descriptions. Each use-case is recognized by its name and description. Use-case description is a natural language description or explanation. This information about the use-case represents field knowledge that is typically available from variants documentation [20], [21], [22].

In our work, use-case description comprises of a small paragraph. Table 2 shows the different releases of mobile media software variants with the use-cases added in each release with its description.

*Table 2: Use-cases of mobile media software variants.*

| # | Use-case | Use-case description |
|---|---|---|
| **Release 1** | | |
| 1 | View album | The client can view an album content on the device memory |
| 2 | Add album | The client can add an album to the mobile media |
| 3 | Delete album | The client can delete an album from the mobile media |
| 4 | Add photo | The client can add a photo in an album presented on the device |
| 5 | Delete photo | The client can delete photo from an album in the device |
| 6 | View photo | The client can view a photo on the device storage |
| 7 | Provide label | The client gives label for the photo and album |
| 8 | Store data | The information of a photo or an album must be saved into the device storage |
| 9 | Remove data | The information of a photo or an album are deleted from the device storage |
| 10 | Retrieve data | The information of a photo or an album are retrieved from the device storage |
| **Release 2** | | |
| * | *Core use-cases* | *All use-cases in release 1* |
| 11 | Count photo | The device keeps count of number of time a photo has been viewed |
| 12 | Edit photo label | The client can edit the existing label of photo and album |
| 13 | View sorted photos | The device sorts the photos by highest viewing frequency |
| **Release 8** | | |
| * | *Core use-cases* | *All use-cases in release 1* |
| 14 | Play video | The client can play the video available in device memory |
| 15 | Capture media | The client can record a video or take a photo using the device camera |

Table 2 shows three releases (*i.e.,* release 1, 2 and 8) of mobile media described by use-cases and their description.

## 3. THE APPROACH OVERVIEW

In this section, we show our goal and core assumptions. Then we introduce our use-case to feature mapping model. Finally, we provide an overview of the process of detecting commonality and variability across use-case diagram variants.

### 3.1 Objective and Basic Assumptions

The main objective of our method is to extract FM for a collection of use-case diagram variants based on the use-cases and their descriptions. Thus, detecting features is the first step towards FM extracting. Common and variable features have been detected from use-case diagrams for a set of product variants. Use-case diagram demonstrates the interactions between the software customer or user and the software processes. Thus, a use-case represents a functional feature in this paper.

### 3.2 Use-case to Feature Mapping Model

The first criteria that allow us to decrease the initial search space is the distinction among common and variable use-cases. Variable use-cases appear in some variants but not in all. We call these use-cases *variable use-cases*. The other use-cases are commons to all variants (*i.e., common use-cases*) (*cf* Figure 3). Each variable feature is implemented by a single use-case.

The second criteria which are utilized to decrease the search space are the variable use-cases common to two or more variants that must be grouped in one cluster. We call the *variable use-cases* are common to two or more variants or those that belong to only one product *Block of Variations* (BV). Then BVs constitute sub search-spaces corresponding to one or many variable use-cases. We rely on this way to decrease the initial search space for the variable use-cases into disjoint clusters (*cf* Figure 3).

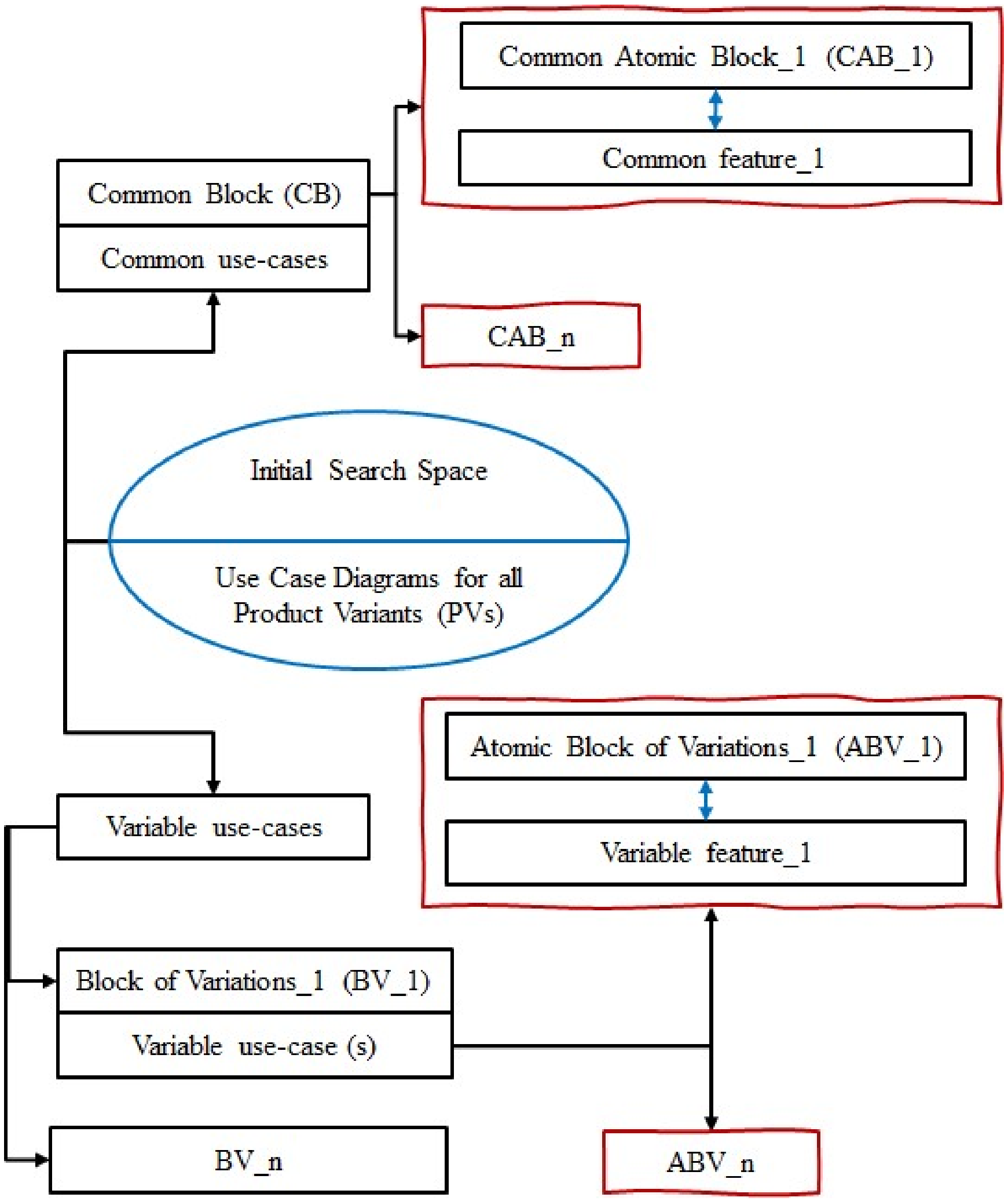

*Figure 3: The initial search space for common and variable use-cases.*

To distinguish between the variable use-cases that appear in the same BV (*cf* Figure 4) we based ourselves on the LSI technique. We rely on the use-case description to detect variable use-case from BVs. This is the third criteria that we used to identify use-case that represent only one variable feature. We call the closest variable use-case for the given description an *Atomic Blocks of Variations* (ABV).

So, each BV is made of a collection of ABVs (*cf* Figure 3) or at least one ABV. Each ABV denotes a single and unique feature.

For a collection of Product Variants (PVs), common use-cases are the use-cases that appear in all PVs. We call these common use-cases that appear in all product variants *common use-cases*. All use-cases that are common to all use-case diagram variants are grouped as one cluster or group called *Common Block* (CB). The CB is composed of several *Common Atomic Blocks* (CABs). To detect the common use-cases, we rely on CB as a search space for this kind of use-cases. All common use-cases which represent the common features appear in the CB as common use-cases. Each CAB represents a single common feature (*cf* Figure 3).

Figure 4 demonstrates an instance of a collection of use-case diagram variants (*three* variants). In Figure 4, the CB is composing the common use-cases that appear in all PVs. In this example, the CB is composing two CABs (*i.e.,* two common features). We can also see the existing BVs, and their composing ABVs (each BV contains at least one ABV). Also, each BV consists of one or more variable use-cases. However, this is an example to present the key concepts which exist in the *use-case to feature mapping model*.

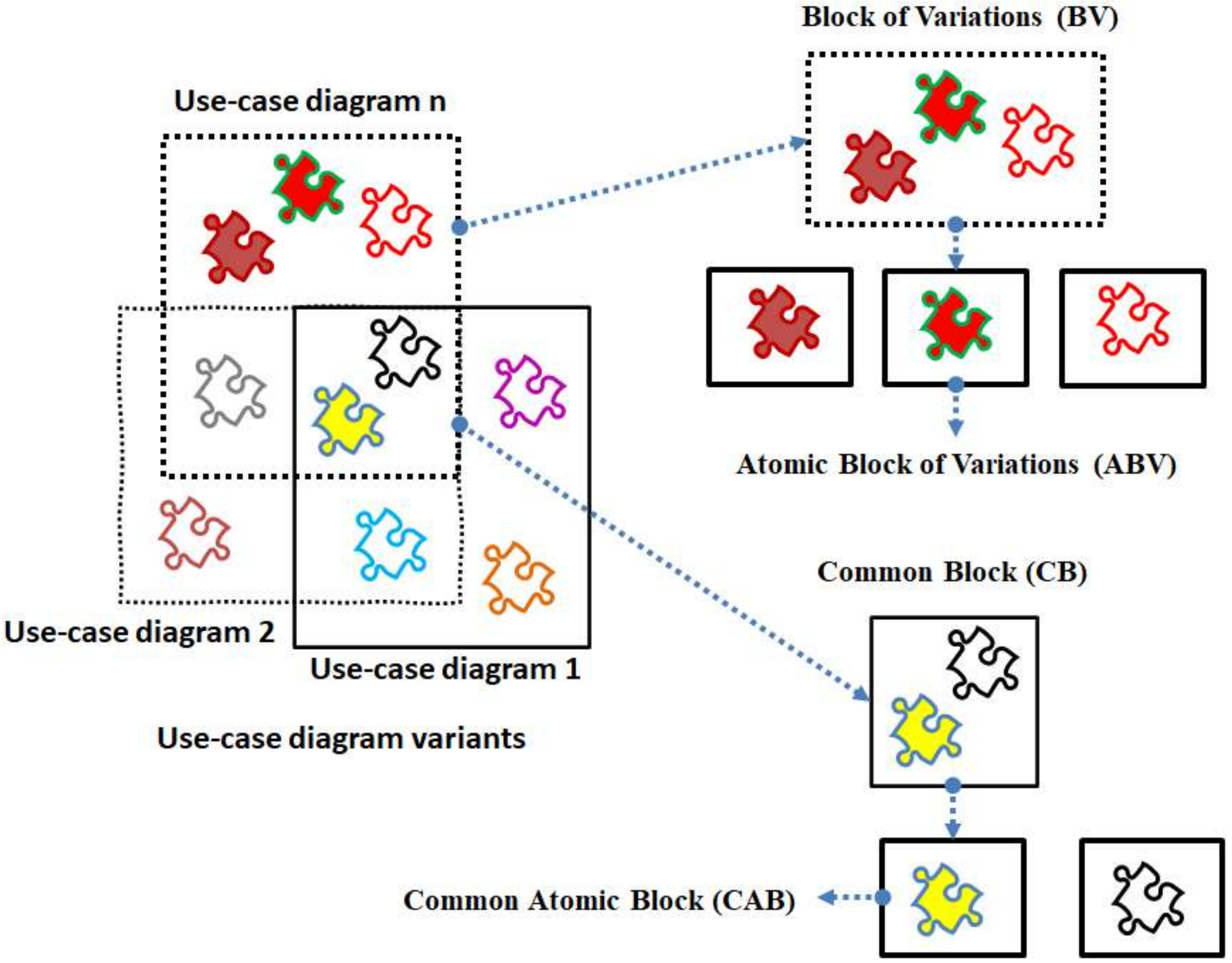

*Figure 4: An illustrative example: CB, CAB, BV, and ABV.*

All concepts defined in the proposed approach for detecting common and variable use-cases (*resp.* features) are illustrated in the use-case to feature mapping model given in Figure 5.

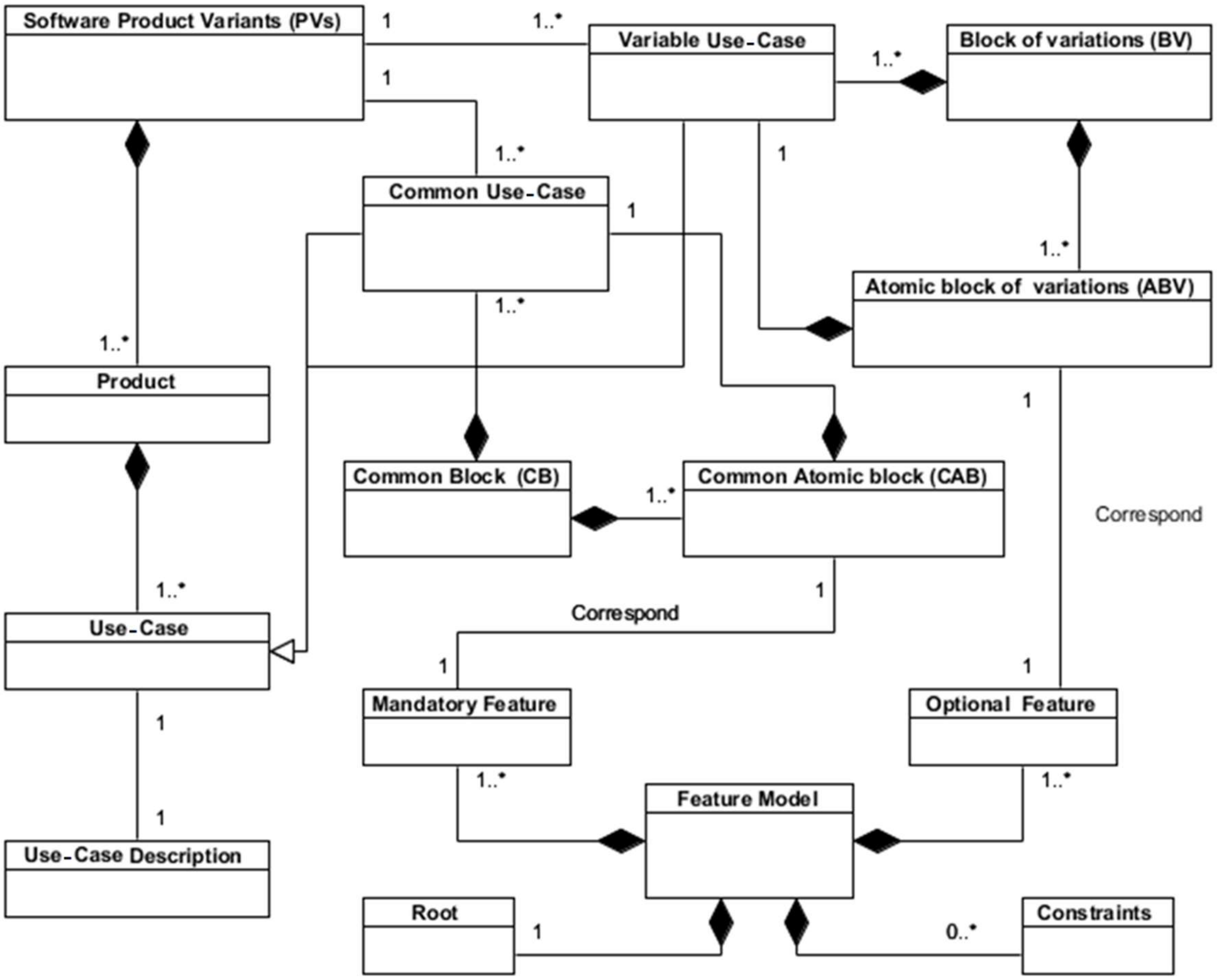

*Figure 5: Use-case to feature mapping model.*

### 3.3 Detecting Commonality and Variability

The mapping model among use-cases and features determines associations among these features and related use-cases for a set of product variants. To determine instances of this model, we rely on a unique process based mainly on FCA and LSI. This process takes as input use-cases and their descriptions of a collection of variants. The main steps of this process are:

- ***Detection of BVs and CB***: FCA is used to detect BVs and CB by reducing the use-cases search space. The *formal context* is defined by the use-case diagram variants which constitute the objects of this context. While all use-cases are the attributes of this context (*cf* Table 3).
- ***Exploring the Lattice of BVs***: The goal of this step is to define an order to search ABVs in the collection of BVs obtained in the *AOC-poset* resulting from the previous step.
- ***Calculate the similarity among use-cases and their description***: Once the CB and BVs are identified, we depend on LSI to define the similarity among use-cases and their description.
- ***Detection of atomic blocks***: The detection of an atomic block is based on the clustering of each use-case with its description. The clustering is based on the similarity measure computed among a collection of use-cases and a collection of use-case descriptions for the given block (CB/BVs). This lexical similarity relationship is used as input for the formal context, where the use-cases consist of its columns, and use-cases descriptions consist of its row.

Figure 6 shows the main steps for detecting commonality and variability in use-case diagram variants.

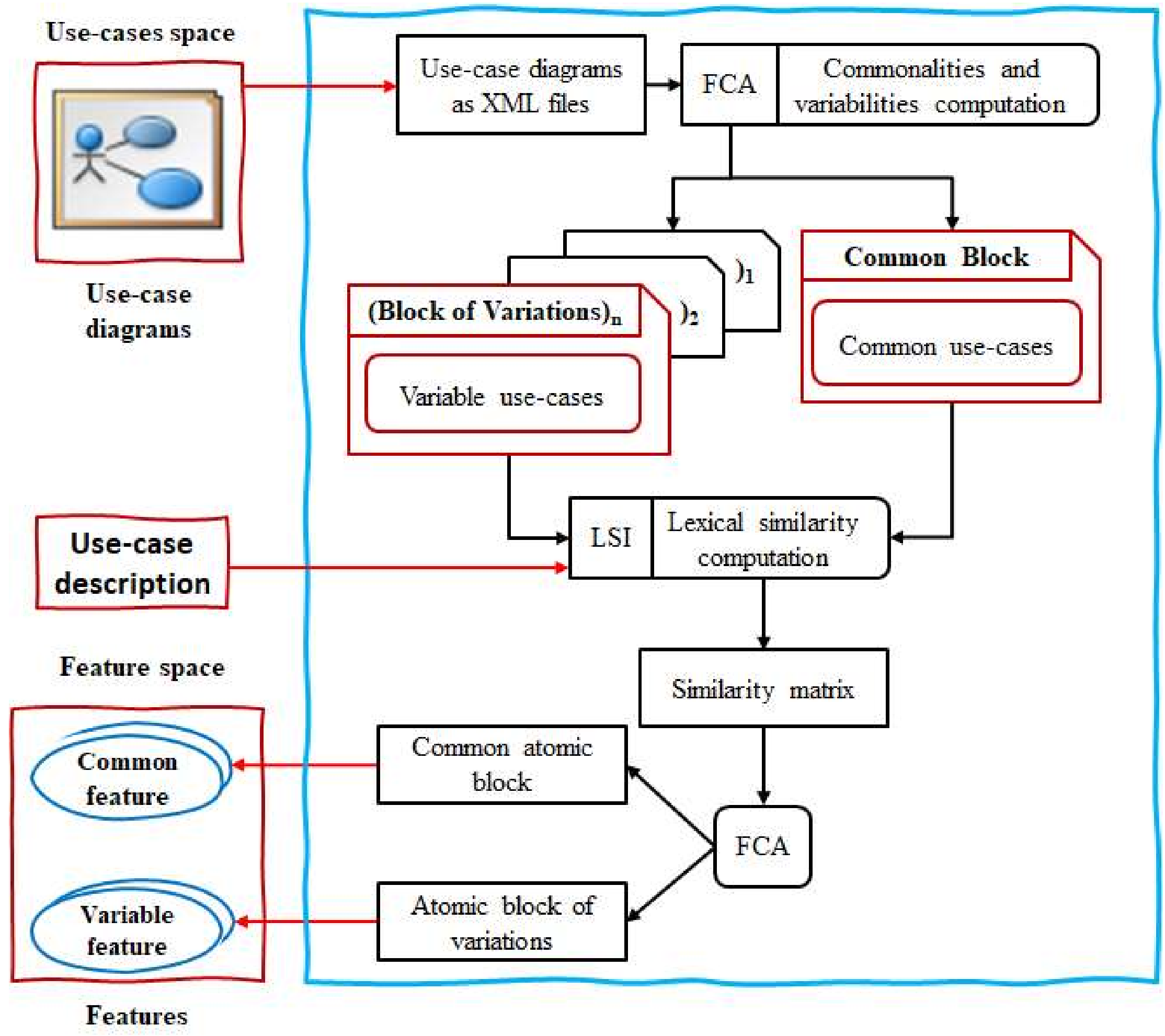

*Figure 6: The process of detecting commonality and variability across use-case diagram variants.*

## 4. THE APPROACH STEP BY STEP

We present in this section, all details about detecting commonality and variability across use-case diagrams, where we explain the FCA role to identify CB and BVs and how we use it again to identify use-cases (*i.e.,* features). Also, we describe how we use LSI to find similarities between use-cases and their description in a particular block (*i.e.,* CB/BVs). All steps are detailed in the following.

### 4.1 Detection of BVs and CB Using FCA

The first phase of the suggested method is the detection of the CB and BVs. The job of these blocks is to be sub-search spaces for detecting common (*resp.* variable) use-cases representing common (*resp.* variable) features. CB and BVs are detected using FCA. The formal context consists of objects (*i.e.,* use-case diagram variants) and attributes (*i.e.,* use-cases). Table 3 shows an instance of formal context. Based on the formal context we obtain an AOC-poset [17]. Figure 7 shows an example of AOC-poset [23]. The *intent* of every concept demonstrates use-cases common to two or more use-case diagram variants. The *extent* of each concept is a group of use-case diagram variants that share these use-cases.

*Table 3: A formal context describing use-case diagram variants of mobile media by their use-cases (partial).*

| | View Album | Delete Album | Add Album | View Photo | Delete Photo | Add Photo | Store Data | Retrieve Data | Provide Label |
|---|---|---|---|---|---|---|---|---|---|
| **re1** | x | x | x | x | x | x | x | x | x |
| **re2** | x | x | x | x | x | x | x | x | x |
| **re3** | x | x | x | x | x | x | x | x | x |
| **re4** | x | x | x | x | x | x | x | x | x |
| **re5** | x | x | x | x | x | x | x | x | x |
| **re6** | x | x | x | x | x | x | x | x | x |
| **re7** | x | x | x | x | x | x | x | x | x |
| **re8** | x | x | x | x | x | x | x | x | x |

| | **Remove Data** | **Error Handling** | **Count Photo** | **Edit Label** | **View Sorted Photos** | **View Favourites** | **Set Favourites** | **Copy Photo** | **........** |
|---|---|---|---|---|---|---|---|---|---|
| **re1** | x | | | | | | | | |
| **re2** | x | x | | | | | | | |
| **re3** | x | x | x | x | x | | | | |
| **re4** | x | x | | | | x | x | | |
| **re5** | x | x | | | | | | x | |
| **re6** | x | x | | | | | | | |
| **re7** | x | x | x | | x | x | x | x | |
| **re8** | x | x | x | x | x | x | x | x | |

As blocks or concepts of AOC-poset in Figure 7 are well-ordered, the intent of the highest concept (T) in the AOC-poset represents use-cases common to all use-case diagram variants. It's the CB. The remaining concepts are BVs. The intent of each of these concepts is a group of use-cases common to a subset of use-case diagram variants but not all variants. The extent of each concept is use-case diagram variants have these use-cases in common.

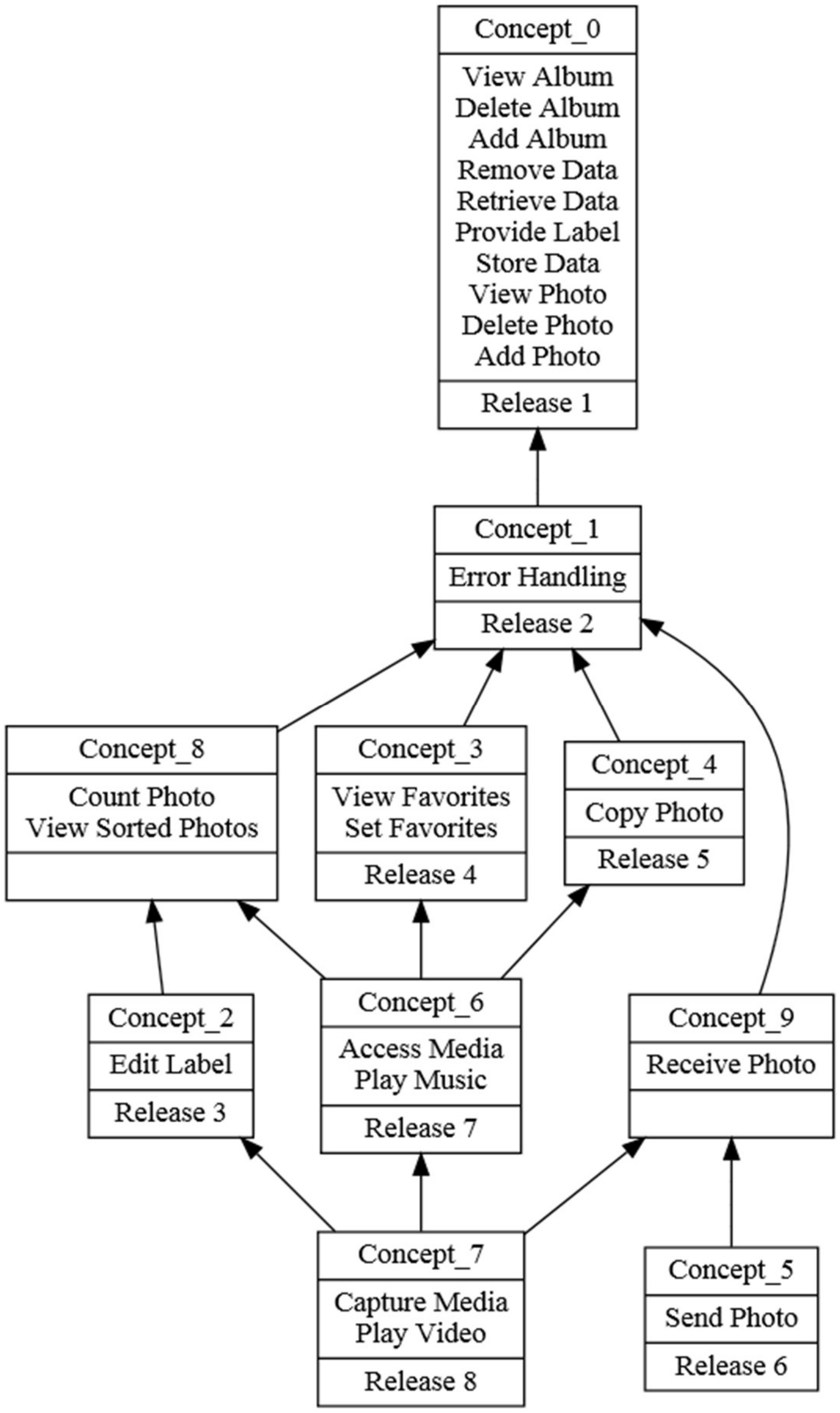

*Figure 7: AOC-poset of formal context in Table 3.*

### 4.2 Detection of Atomic Blocks (or features)

CB and BVs are sub-search spaces for common and variable features respectively. Concepts that are obtained in the AOC-poset were generated in the previous step, and whose intents consist of BVs, are ordered. Thus, to optimize the search process for the ABVs in the obtained collection of BVs we explore this AOC-poset starting from the most specific spaces to the most general ones. Atomic blocks are detected based on the computation of the similarity value among the use-cases and their description in a given BV. These similarity values are obtained by applying LSI. Atomic blocks are generated by clustering the use-cases based on the similarity between these use-cases and their description using FCA. The search strategy that we follow in this paper starts from the bottom concepts to the top one, where BV in the bottom contains fewer use-cases in its intent compared to the next level. Atomic blocks are groups of the most similar use-cases and their descriptions created with FCA as described in the following.

#### 4.2.1 Exploring BVs to Detect ABVs

We need to optimize the search process for the variable features. Optimization means that search is gradually done from the smaller BV to the larger ones. However, this means that compared to the concepts of the AOC-poset, the search is performed from the most specific concepts (*i.e.,* ⊥) to the general concept (*i.e.,* ⊤). Thus, Thus, the order of exploration of the AOC-poset is from the more specific concepts to the more general concept. This process aims to detect variable features from the smallest search space. For CB, there is only one block in the AOC-poset. Thus, no need for exploring the AOC-poset for CB (*i.e.,* top concept).

#### 4.2.2 Measuring Similarity Between Use-cases and their Description Based on LSI

Use-cases of BV or CB are respectively representing variable and common features. We detect use-case (*i.e.,* feature) from BV (or CB) based on the measurement of lexical similarity among use-cases and their description from every block. This lexical similarity measure has been determined by using LSI. To calculate the similarity among use-cases and their description in CB or BVs, the authors proceed in 3 steps: **1)** creating LSI corpus, **2)** creating term-document matrix and term-query matrix for each BV/CB, and **3)** creating cosine similarity matrix as described in the following.

##### 4.2.2.1 Creating LSI Corpus

To utilize LSI in our work, we must create a corpus that denotes a set of documents and queries. In our work, every use-case in the BV (*resp.* CB) denotes a single document and its description exemplify a query. All information in a document and a query must be normalized to become appropriate as input of LSI. All documents and queries must be divided into terms and execute word-stemming [24].

##### 4.2.2.2 Creating Term-Document Matrix and Term-Query Matrix for CB and BV

The term-document matrix in the LSI is described by $t \times u$ matrix [25]. In our case $t$ is the number of terms utilized in a normalized document relating to a use-case and $u$ is the number of use-cases in CB (or BV). In the same manner, a term-query matrix is represented by $t \times d$ matrix where $t$ is the number of terms and $d$ number of use-cases descriptions. In our method, LSI utilizes use-case description as a query to recover related use-case.

##### 4.2.2.3 Creating Similarity Matrix

The similarity among use-cases and their descriptions in BV or CB is defined by the cosine similarity matrix. In similarity, matrix columns denote use-cases vectors (*i.e.,* documents) and rows denote use-cases description vectors (*i.e.,* queries). The similarity is computed based on cosine similarity [25]. In our method, we consider the best utilized threshold for cosine similarity that equals *0.70* [25]. Thus, only the use-case name and use-case description that have similarity value larger than or equal to the threshold (*i.e.,* >=70) are considered as similar. Table 4 reveals the cosine similarity matrix of Concept_0 (*i.e.,* common block) in Figure 7.

*Table 4: The similarity matrix of Concept_0 in Figure 7 (partial).*

| **Use-case** / **Description** | Add album | Add photo | Delete album | … |
|---|---|---|---|---|
| Add album des. | **1** | 0.52 | 0.1 | |
| Add photo des. | 0.55 | **0.99** | -0.19 | |
| Delete album des. | 0.1 | -0.2 | **1** | |
| Delete photo des. | -0.19 | 0.38 | 0.55 | … |
| Provide label des. | -4.77 | 0 | -4.77 | |
| Remove data des. | -5.24 | 0 | -5.24 | |
| Retrieve data des. | -5.24 | 0 | -5.24 | |
| Store data des. | -5.24 | 0 | -5.24 | |
| View album des. | 0.18 | -0.36 | 0.18 | |
| View photo des. | -0.18 | 0.35 | -0.18 | |

#### 4.2.3 Detecting Atomic Blocks Based on FCA

We use FCA to detect the atomic feature (*i.e.,* use-case) from each block (BVs and CB). We use block contents (*i.e.,* use-cases) with additional information (*i.e.,* use-case descriptions) to detect features based on the similarity between use-case name and use-case description. This similarity is based on LSI. The authors consider each use-case as a feature. A Formal context consists of objects (*i.e.,* use-case descriptions) and attributes (*i.e.,* use-cases). Table 5 displays the formal context achieved by converting the similarity matrix (*cf* Table 4) relating to *Concept_0* from Figure 7 (*i.e.,* CB).

*Table 5: Formal context of Concept_0 (partial).*

| **Use-case Description** | Add album | Add photo | Delete album | … |
|---|---|---|---|---|
| Add album des. | x | | | |
| Add photo des. | | x | | |
| Delete album des. | | | x | |
| Delete photo des. | … | | | |
| Provide label des. | … | | | |
| Remove data des. | … | | | |
| Retrieve data des. | … | | | |
| Store data des. | … | | | |
| View album des. | … | | | |
| View photo des. | … | | | |

The *binary relation* that is used in this context is a similarity among use-cases and their description according to LSI, so we used similarity matrix (*cf* Table 4) (*i.e.,* LSI findings) as input for the FCA to gather together the use-case and its description by using the lexical similarity. The AOC-poset in Figure 8 shows ten CAB (*i.e.,* 10 common features) detected from CB (*cf* Concept_0 in Figure 7). Each CAB represents one and only one common feature (*i.e.,* mandatory feature).

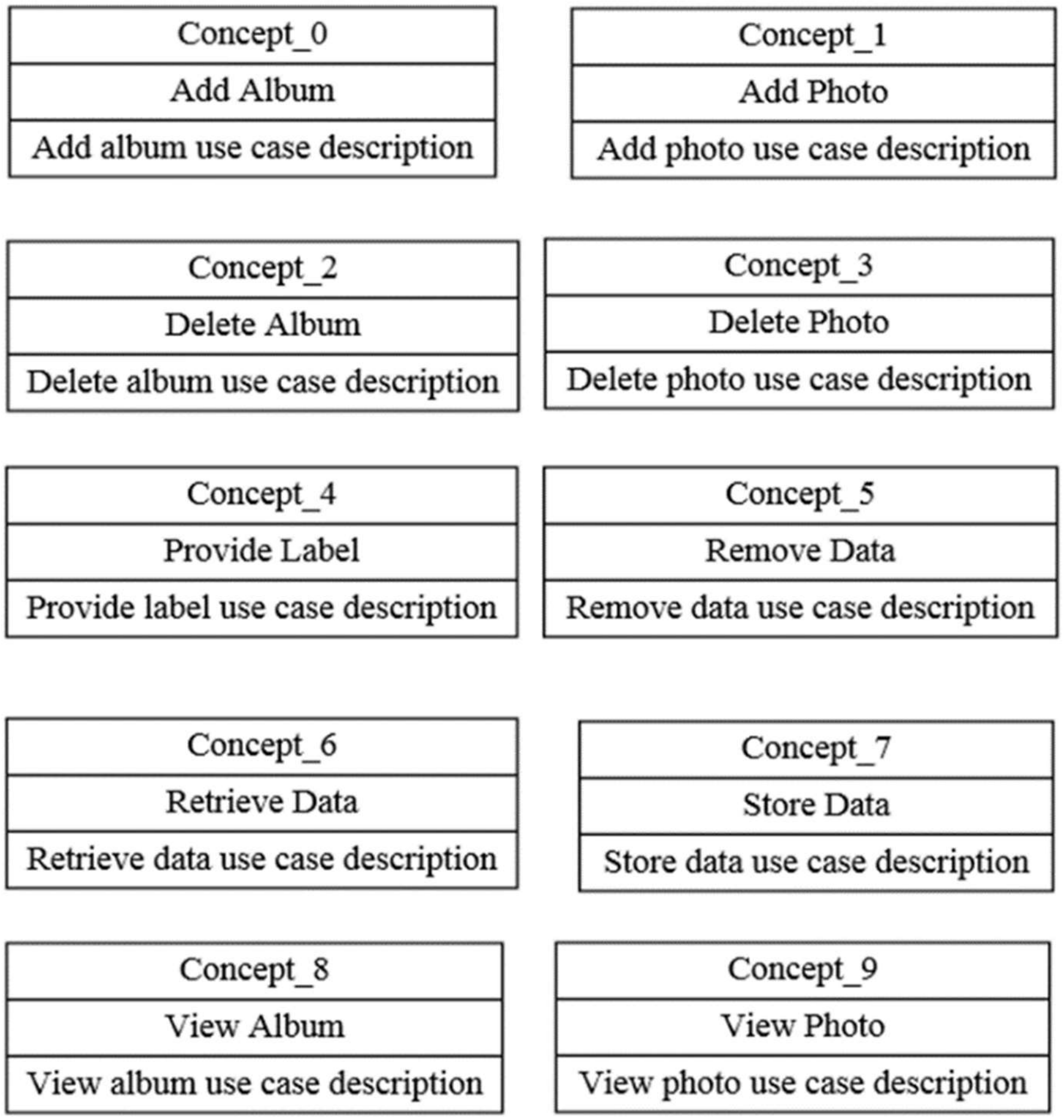

*Figure 8: AOC-poset shows ten CAB (i.e., common features) detected from CB (i.e., Concept_0 in Figure 7).*

The result in Figure 8 is an AOC-poset to which the extent of each concept represents a use-case description. However, the intent of each concept is the use-case name. The extents of these concepts are use-case descriptions that are directly linking to its intent based on the similarity link.

Figure 9 shows the detected variable features (*i.e.,* optional features) from Concept_3 (*i.e.,* BV) in Figure 7. Where both features are variable features detected from the same BV.

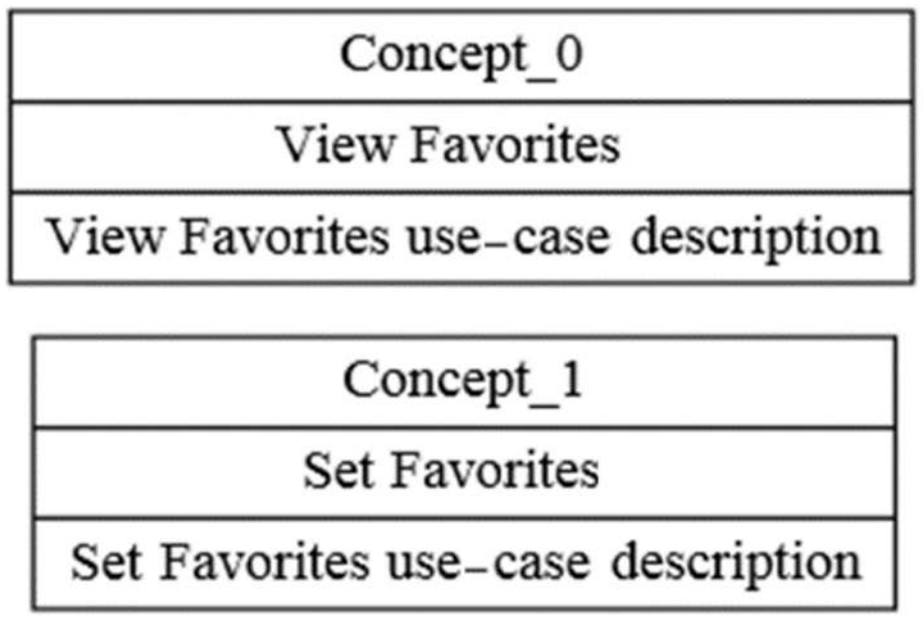

*Figure 9: AOC-poset shows two ABVs (i.e., variable features) detected from Concept_3 (i.e., BV) in Figure 7.*

## 5. EXPERIMENTATION

To justify our approach, we have experimented with it on use-case diagram variants of mobile media [6]. We used eight variants from mobile media. Mobile media variants are well documented, and their FM is available. However, this has allowed us to compare our obtained results with the known feature of these variants. In addition, mobile media manipulates multimedia (*i.e.,* photos, music, and video) on mobile devices. Mobile media suffered seven evolution scenarios (*cf* Section 2), which produced eight releases of this family. The scenarios contain several kinds of changes including common and variable features [16]. Table 6 presents the achieved results for use-case diagram variants of mobile media. In this table, we gave names to atomic blocks by using the use-case names.

*Table 6: The detected commonality and variability from use-case diagram variants of mobile media case study.*

| # | Feature name | Feature type | | F-Meas. |
|---|---|---|---|---|
| | | Comm. | Vari. | |
| 1 | View album | X | | 100% |
| 2 | Add album | X | | 100% |
| 3 | Delete album | X | | 100% |
| 4 | Add photo | X | | 100% |
| 5 | Delete photo | X | | 100% |
| 6 | View photo | X | | 100% |
| 7 | Provide label | X | | 100% |
| 8 | Store data | X | | 100% |
| 9 | Remove data | X | | 100% |
| 10 | Retrieve data | X | | 100% |
| 11 | Error handling | | X | 100% |
| 12 | View sorted photos | | X | 100% |
| 13 | Edit label | | X | 100% |
| 14 | Count photo | | X | 100% |
| 15 | Set favorites | | X | 100% |
| 16 | View favorites | | X | 100% |
| 17 | Copy photo | | X | 100% |
| 18 | Send photo | | X | 100% |
| 19 | Receive photo | | X | 100% |
| 20 | Play music | | X | 100% |
| 21 | Access media | | X | 100% |
| 22 | Play video | | X | 100% |
| 23 | Capture media | | X | 100% |

The *results* show that the F-Measure metric seems to be high for all detected features (*i.e.,* 100%). This implies that all features are detected from use-case diagram variants of mobile media. The F-Measure metric shows the efficiency of our approach, where it is based on precision and recall metrics. A high F-Measure means an ideal approach.

We verify our method on the mobile media case study. The method successfully presents to the software engineer variability and commonality of use-case diagrams in terms of features. This reveals that it is a unique method and that it can be applied to any kind of Unified Modeling Language (UML) diagrams. Figure 10 displays the detected FM from use-case diagram variants of mobile media. The detected FM consists of optional and mandatory features based on the obtained results.

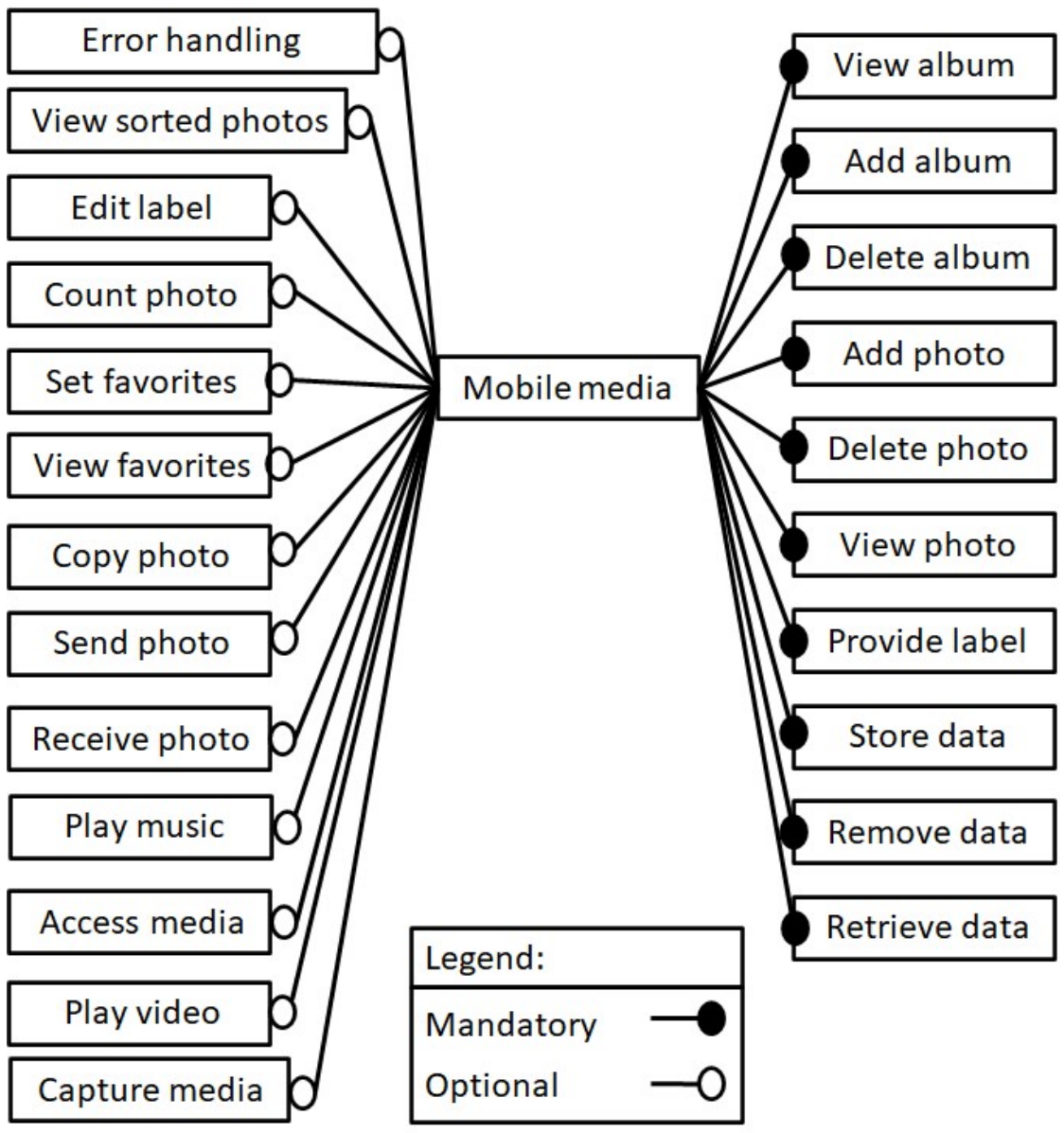

*Figure 10: FM of use-case diagram variants of mobile media.*

The case study that we used in this work is well-documented. This study also provides the real variability and commonality in terms of features. Thus, we manually assessed the detected features with those given in FM of the mobile media. The assessment indicates that we get a complete matching among commonality and variability detected by our method and those shown in the FM of the mobile media.

The *threat to the validity* of our method is that software designers may don't use the same vocabularies to label use-cases across diagram variants. This implies that lexical similarity maybe not be reliable to detect common and variable use-cases. Also, we assumed in this paper that every use-case represents a functional feature without considering other elements such as actors and relations among use-cases.

## 6. RELATED WORK

In this section, we describe the closest studies, which are relevant to the detection of commonality and variability from software model or diagram variants.

Rubin *et al.* [26] suggested an approach to compare a collection of UML state charts to refactor them as a SPL. The authors aim to combine input variants into a general model and not at detecting commonality and variability among software model variants.

Martinez *et al.* [27] suggested an approach called model variants comparison. Their approach compares a collection of model variants and extracted commonality and variability among them. In their work, each extracted feature involves a collection of atomic model components. However, their approach visualizes the extracted features by using a graphical depiction where common (*resp.* variable) features are offered to software engineers.

Ziadi *et al.* [28] suggested an approach to analyzing the code of software variants via the usage of UML class diagrams to identify commonality and variability among software variants. This approach is not generic, and the authors do not provide a visualization of commonality and variability. They suggested a method for feature location in the code of a collection of product variants. Their method gathers all common features as a single common feature under the title *base feature* without distinguishing between the variable features that exist in the same block. Our approach relies on the use-case diagram variants to detect both common and variable features. FCA is used to extract commonalities and variabilities from use-case diagram variants, and we differentiate among the common and variable features by employing LSI and FCA.

Al-Msie'deen *et al.* [29], [30] presented a method to obtain commonality and variability from the code of a collection of variants by using the FCA. Their work visualizes the extracted commonality and variability via the mined AOC-poset where commonality (*resp.* variability) across software variants source code is offered to software developers. Al-Msie'deen and Blasi [31] detected the evolution scenario across the code of two product variants using FCA. Al-Msie'deen *et al.* [32] presented a method to display the software variability based on different software artifacts using FCA.

Current approaches in the field of SPL reengineering compare a collection of diagram variants at the same time. The representation and visualization of commonality and variability are not considered because their key purpose involves refactoring diagram variants to a SPL without involving a software engineer in the process. We say that the software engineer should take a role in the analysis of the detected commonality and variability information for SPL approval. In this work, we suggest a method that lets comparing a collection of use-case diagrams at the same time to detect commonality and variability among them. Our method offers a graphical visualization of the comparison result using FCA.

## 7. CONCLUSION

In this paper, an approach has been proposed using FCA and LSI for detecting commonality and variability from use-case variants. The use-case diagram variants comparison method was proposed as a facilitator to detect and analyze commonality and variability information in a collection of diagrams. FCA is used to obtain the commonalities and variabilities from a set of use-case diagram variants. LSI is utilized with FCA to detect atomic block that represents a unique feature from CB and BVs by using the lexical similarity among the use-cases and their description. We validated the suggested approach on a mobile media case study. In future work, the plan is to utilize the detected common and variable features to construct FM [33], [34], [35]. Also, the suggested approach can be employed on an industrial scenario that deals with large use-case diagram variants. Finally, we plan to apply the word cloud technique [36], [37] to name the detected blocks based on the words extracted from atomic blocks.